\documentclass[final]{cvpr}

\usepackage{amsmath}
\usepackage{subcaption}
\usepackage{amsfonts}
\makeatletter
\@namedef{ver@everyshi.sty}{}
\makeatother
\usepackage{tikz}
\usepackage{contour}

\usepackage{mypkg}

\usepackage[pagebackref=true,breaklinks=true,colorlinks,bookmarks=false]{hyperref}

\def\cvprPaperID{} 
\def\confYear{}

\begin{document}

\title{Rectangle-based Approximation for Rendering \\Glossy Interreflections}

\author{Chunbiao Guo}

\maketitle
\thispagestyle{empty}
\pagestyle{empty}

\begin{abstract}
This study introduces an approximation for rendering one bounce glossy interreflection in real time. The solution is based on the most representative point (MRP) and extends to a sampling disk near the MRP. Our algorithm represents geometry as rectangle proxies and specular reflections using a spherical Gaussian. The reflected radiance from the disk was efficiently approximated by selecting a representative attenuation axis in the sampling disk. We provide an efficient approximation of the glossy interreflection and can efficiently perform the approximation at runtime. Our method uses forward rendering (without using GBuffer), which is more suitable for platforms that favor forward rendering, such as mobile applications and virtual reality.
\end{abstract}

%
\section{Introduction}
Real-time glossy interreflections are important visual cues for synthetic images. These interreflections are important for interactive virtual environments, such as virtual reality (VR) applications or video games. The real-time rendering of glossy interreflections is a challenging task. The difficulty resides in the evaluation of a two-dimensional specular radiance integral for which no practical solution exists, except for expensive Monte Carlo-based sampling techniques. Most compelling solutions are found using the most representative point (MRP) approaches, thereby reducing the shading integration problem to a cheap point-lighting calculation. However, these methods fail to preserve the specular highlight shape of underlying bidirectional reflectance distribution functions (BRDFs). Moreover, finding an approximation for real-time performance remains a problem. Despite advances in graphics hardware and corresponding algorithms, it is currently impossible to render accurate glossy interreflections.

We classify real-time global illumination(GI) algorithms into precomputation and post-processing methods. The former uses precomputed data to accelerate real-time computation. Screenspace methods solve GI problems by sparsely sampling the space, which is simple to implement but requires additional filtering to remove noise, and most of them can only handle diffuse interreflection. In some platforms favoring forward rendering, such as VR applications and mobile platforms, the screenspace method is not applicable. In addition, in some platforms that lack computation performance, such as computer games running on mobile devices, there are currently no methods to approximate such effects. 

This study presents a new forward rendering method for rendering glossy interreflections at real-time frame rates. Our method is based on a rectangle-based structure that is used to approximate the origin geometry. We thereafter used a spherical Gaussian (SG) to approximate the reflected lighting in a rectangle for a given shading point. Our algorithm can promptly approximate the integral of reflection lighting on a rectangle in real time. 

In conclusion, the contributions of this study are as follows:
\begin{itemize}
	\item We provide a functional approximation of the glossy interreflection, and we can efficiently perform the approximation at run time.
	\item We develop a novel spherical Gaussian representation of glossy reflection.
	\item We investigate the real-time GI algorithm using forward rendering.
\end{itemize}
	
\section{Background}
\subsection{Related work}
We first review some related works and introduce the necessary background of the SGs representation. 

\textbf{Precompute methods.} Precomputed radiance transfer(PRT) \cite{Sloan2002} precomputes linear operators that map the light transport equation into indirect radiance or irradiance sampled over the scene surfaces. The precomputed functions are stored using basis functions, such as spherical harmonics (SHs) \cite{Ramamoorthi2001,WangRui2009}, wavelet \cite{Ng2003}, and SG \cite{Tsai2006}. This method assumes that the scene geometry is fixed and only the light and camera change. The main idea is to project the light transport function into the frequency domain using a spherical basis and thereafter perform real-time multiplication in the frequency domain rather than time-consuming convolution in the space domain. However, this method cannot handle dynamic objects and can only represent low-frequency features. Photon mapping \cite{Hachisuka2008, Hachisuka2009} is widely used in video games: First, photons are emitted from a light source and recorded on the geometric surface of the scene and subsequently organized in a spatial data structure. Thereafter, a second pass starting from the surfel of light-field texture while gathering nearby photons to estimate indirect illumination. Finally, these light-field textures were used for real-time rendering.

\textbf{Postprocessing methods.} Some methods do not require precomputation to calculate GI. A reflective shadow map \cite{Dachsbacher2005,Dachsbacher2006} considers each pixel in the colored shadow map an indirect light source (virtual point light \cite{Keller1997,Walter2005,Walter2006}). Light propagation volumes (LPVs) \cite{Kaplanyan2010} store lighting information from light in a 3D grid. Every grid represents the distribution of the indirect light in the scene. Light is propagated from the 3D grid to nearby grids. The drawback of LPV is that it can only use one distant light source. Screenspace directional occlusion (SSDO) \cite{Ritschel2009GI} accounts for the direction of the incoming light and includes one bounce of indirect illumination. Because of the sparse sampling of SSDO, sophisticated nonlinear filters are required to remove noise (e.g., \cite{Kontkanen2004}). SSDO has been widely adopted in the video game industry because it relies on GBuffer; therefore, it cannot be used well under certain conditions such as VR and some mobile platforms and can only handle diffuse indirect illuminations. Our method is based on forward rendering, without the use of GBuffer. In addition, almost all of these methods reuse screen-space illumination generated by the camera to compute the next bounces of light transport. This approximation works well if the scene is diffuse, or the next bounces of light transport are insensitive to directions. However, for glossy reflections, such an approximation is no longer correct. Our method considers the BRDF direction of the glossy reflections.

\textbf{Geometric approximations.} Some methods use geometric approximations to perform faster integration. Point-based models \cite{Per2008} store surface elements of shape in a point cloud file and are organized into an octree hierarchy; thereafter, the illumination from the surface point in each octree node is approximated using SHs. Voxel cone tracing (VXGI) \cite{Crassin2011,Ritschel2009} approximates one bounce of light in a dynamic scene, a voxel representation is created once for static objects and every frame for dynamic objects. Each voxel stores the amount of light that the geometry emits in all directions, and the irradiance is calculated using cone tracing. The results of VXGI are satisfactory, but cone tracing is time-consuming. In this study, we use a rectangle to accelerate integration. 

\textbf{Most representative point.} In addition to geometric approximations, Picott \cite{Picott1992} first introduced MRP approaches to alleviate the costly sampling computation by identifying a representative point in the light area that contributes the most to the illumination. Karis \cite{Karis2013} used a modification of the specular distribution to ensure energy conservation, making the result better match the intensity highlight of specular microfacet models. We simply extend these methods by selecting a specular peak point as a representative point for specular interreflection, which builds on the property that the reflected light path half vector coincides with the surface normal at the specular peak point.

\textbf{SG lighting.} Compared to SHs, SGs \cite{Wang2009} are more compact for all frequency effects. Since several fundamental rendering operations such as integral, product, and product integral have closed-form solutions in SG, SG can be used to represent the different parts of the rendering equation to approximate the illumination integral. The Gaussian approximation for the reflectance of several microfacet BRDF models can be derived in a closed form. Since a normalized SG is equivalent to the von Mises--Fisher distribution, several SGs can be merged into a single SG in an analytic way. Wang et al. \cite{Wang2009} developed an SG approximation for spatially varying BRDFs (SVBRDFs), allowing for real-time all-frequency rendering. Iwasaki et al. \cite{Iwasaki2012} introduced an integral SG and reduced the need for precomputed data. Anisotropic SGs (ASGs) \cite{Xu2013} have been used to represent anisotropic lighting and BRDFs. Pettineo \cite{Pettineo2016} used SGs to encode light-field textures in the videogame The Order 1886. Their experiment demonstrated that with 12 SGs with fixed direction and sharpness (i.e., 36 floats), they can better represent the original light field than a 3-band SH representation (24 floats). Xu et al.\cite{Xu2014} used the SG to approximate local interreflections. They integrated the projection of a triangle over a sphere as an SG, and the performance was linked to the geometry tessellation. However, their method cannot achieve real-time performance because they usually need to integrate several triangles, which severely reduces the performance. Our work approximates the geometry as rectangles first and subsequently uses this approximation to reduce the calculation cost; we only need to calculate a few rectangles for each pixel.

\begin{figure*}
	\begin{subfigure}{0.24\textwidth}
		\includegraphics[trim=50 50 50 50, clip, width=1\textwidth]{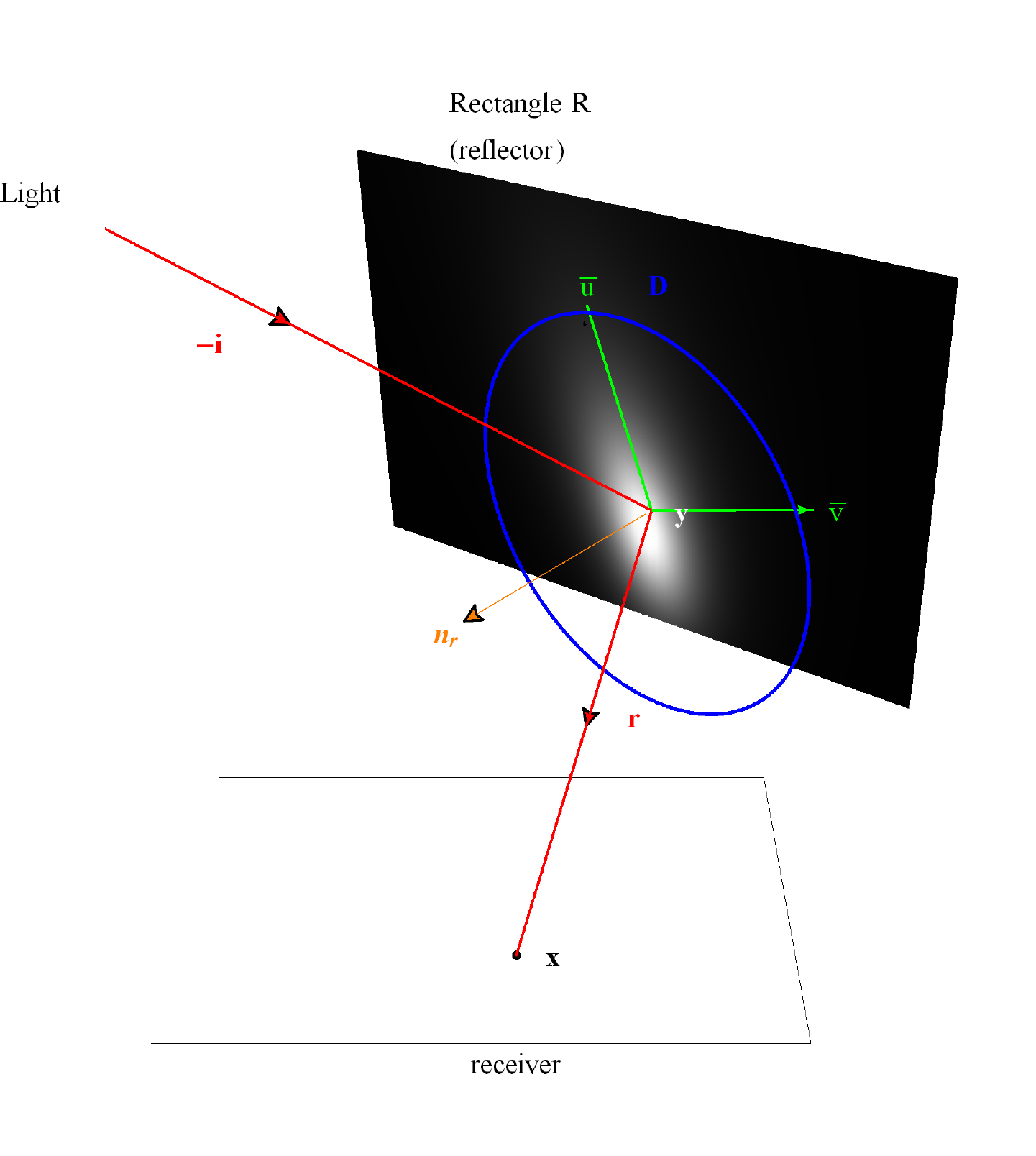}
		\caption{}
		\label{fig:overview_a}
	\end{subfigure}
	\begin{subfigure}{0.24\textwidth}
		\includegraphics[trim=40 40 40 40, clip, width=1\textwidth]{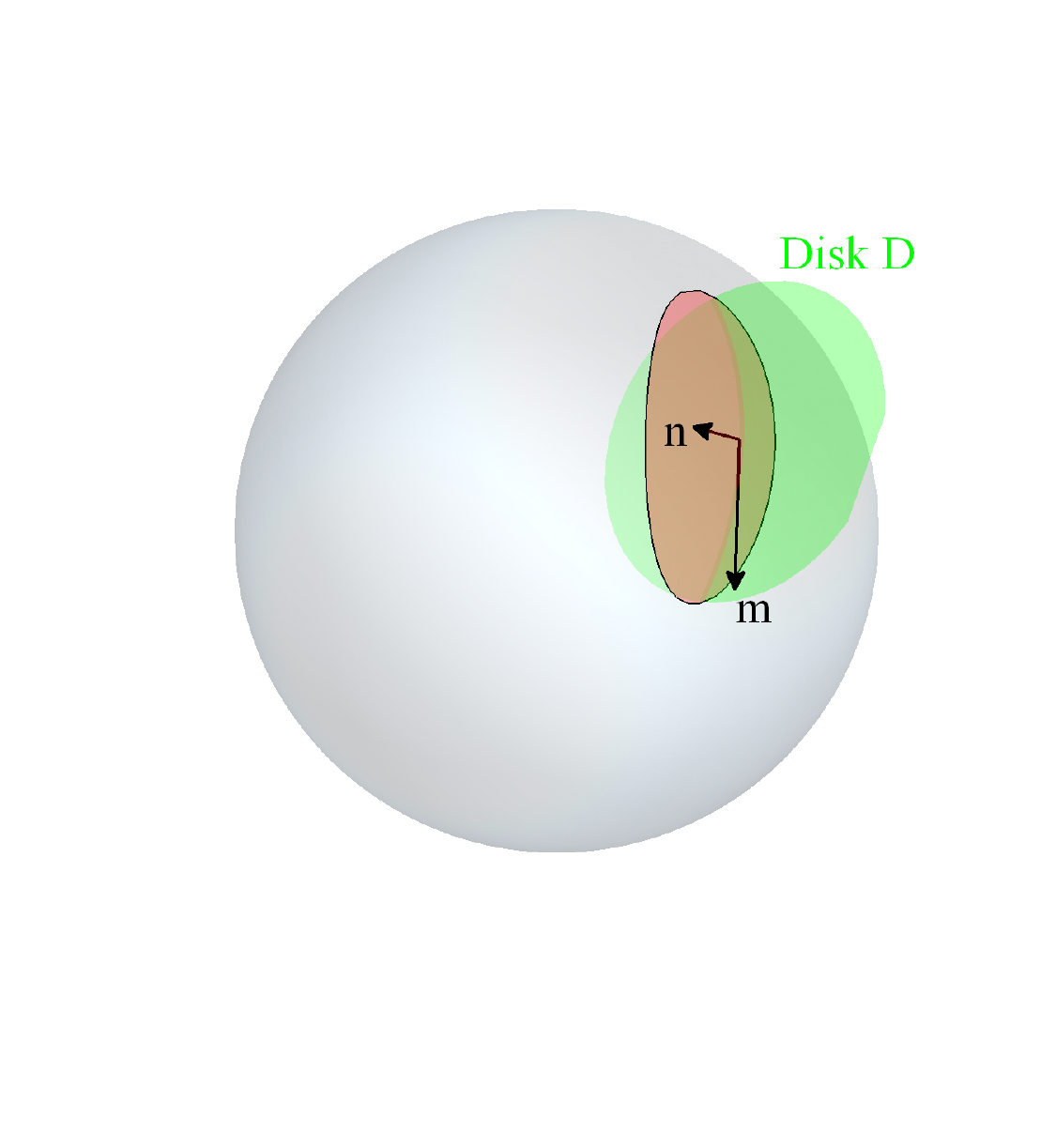}
		\caption{}
		\label{fig:overview_b}
	\end{subfigure}
	\begin{subfigure}{0.24\textwidth}
		\centering
		\includegraphics[trim=40 40 40 40, clip, width=1\textwidth]{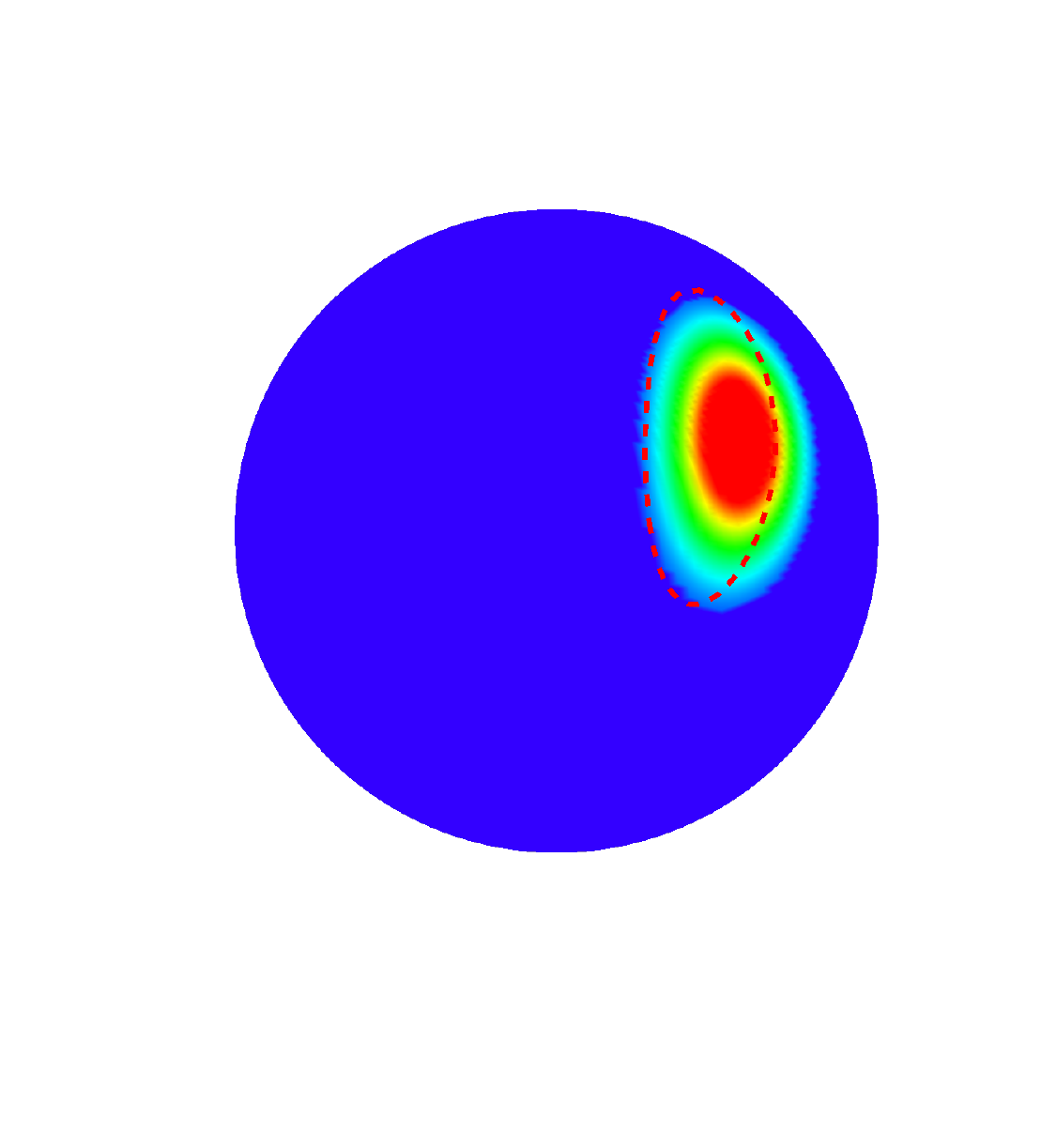}
		\caption{}
		\label{fig:overview_c}
	\end{subfigure}
	\begin{subfigure}{0.24\textwidth}
		\centering
		\includegraphics[trim=0 0 0 120, clip, width=1\textwidth]{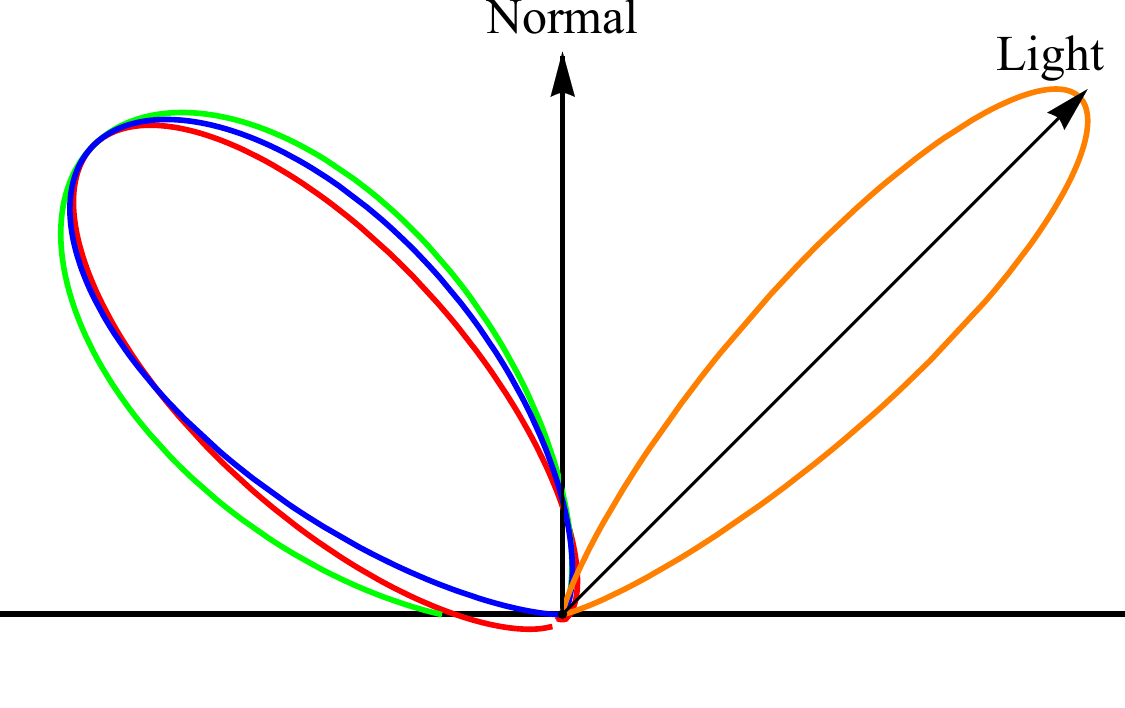}
		\caption{}
		\label{fig:overview_d}
	\end{subfigure}
	\caption{Overview}
	\medskip 
	\small
	\textbf{(a)}. A specular peak point $\peakPt$ is selected on the reflector plane; $\disk$ is then selected to gather the one-bounce reflected radiance from rectangle $\rectangle$ to shading point $\recvPt$, and we also observed that the reflected radiance can be represented with a representative attenuation axis $\uAxis$. \textbf{(b)}. The integrated reflected radiance from the disk is treated as the disk (green) and projected onto a sphere; the projected area (red) is an ellipse-like shape, and the ${\asgBitangent, \asgTangent}$ axes of the ellipse correspond to the ${\uAxis}$ axes of the disk. \textbf{(c)}. The projected area (red dotted lines) was approximated using an ASG. \textbf{(c)}. The NDF (red lobe) is approximated with an SG (green lobe) then warped from the half-vector domain to the lighting domain (blue lobe), and subsequently convolved with ASG light (orange lobe).
	\label{fig:overview}
\end{figure*}

\subsection{Spherical Gaussians}
An isotropic SG (SG) is a type of spherical function for the direction $\mathbf{v}$ as follows:
\begin{equation}
	G_{i}(\mathbf{v}; \mathbf{p}, \nu, a) = a \cdot \, e^{\nu (\mathbf{v} \cdot \mathbf{p} - 1)},
\end{equation}
where $\mathbf{v} \in \mathbb{S}^2$ denotes the function input, $\mathbf{p} \in \mathbb{S}^2$ is the lobe axis, $\nu \in \mathbb{R}_+$ is the lobe sharpness, and $a \in \mathbb{R}_+^n$ is the lobe amplitude.

An ASG behaves similar to SGs but is anisotropic rather than isotropic:
\begin{equation}\label{eq:asg} 
	G_{a}(\mathbf{v}; [\mathbf{z},\mathbf{x},\mathbf{y}], [\lambda,\mu], c) = c \cdot \mathbf{S}(\mathbf{v};\mathbf{z}) \cdot e^{-\lambda (\mathbf{v} \cdot \mathbf{x})^2-\mu (\mathbf{v} \cdot \mathbf{y})^2}.
\end{equation}
where $\mathbf{z} \mathbf{x},\mathbf{y}$ denote the lobe, tangent, and bi-tangent axes, respectively, and $\lambda$ and $\mu$ are the bandwidths for the $\mathbf{x}$- and $\mathbf{y}$-axes, respectively, satisfying $\lambda,\mu > 0$. The term $c$ denotes the lobe amplitude; the smooth term is defined as follows: $\mathbf{S}(\mathbf{v};\mathbf{z})=\max(\mathbf{v} \cdot \mathbf{z},0)$.
	
\section{Rendering}
\begin{figure}[!b]
	\includegraphics[trim=50 50 50 50, clip,width=\linewidth]{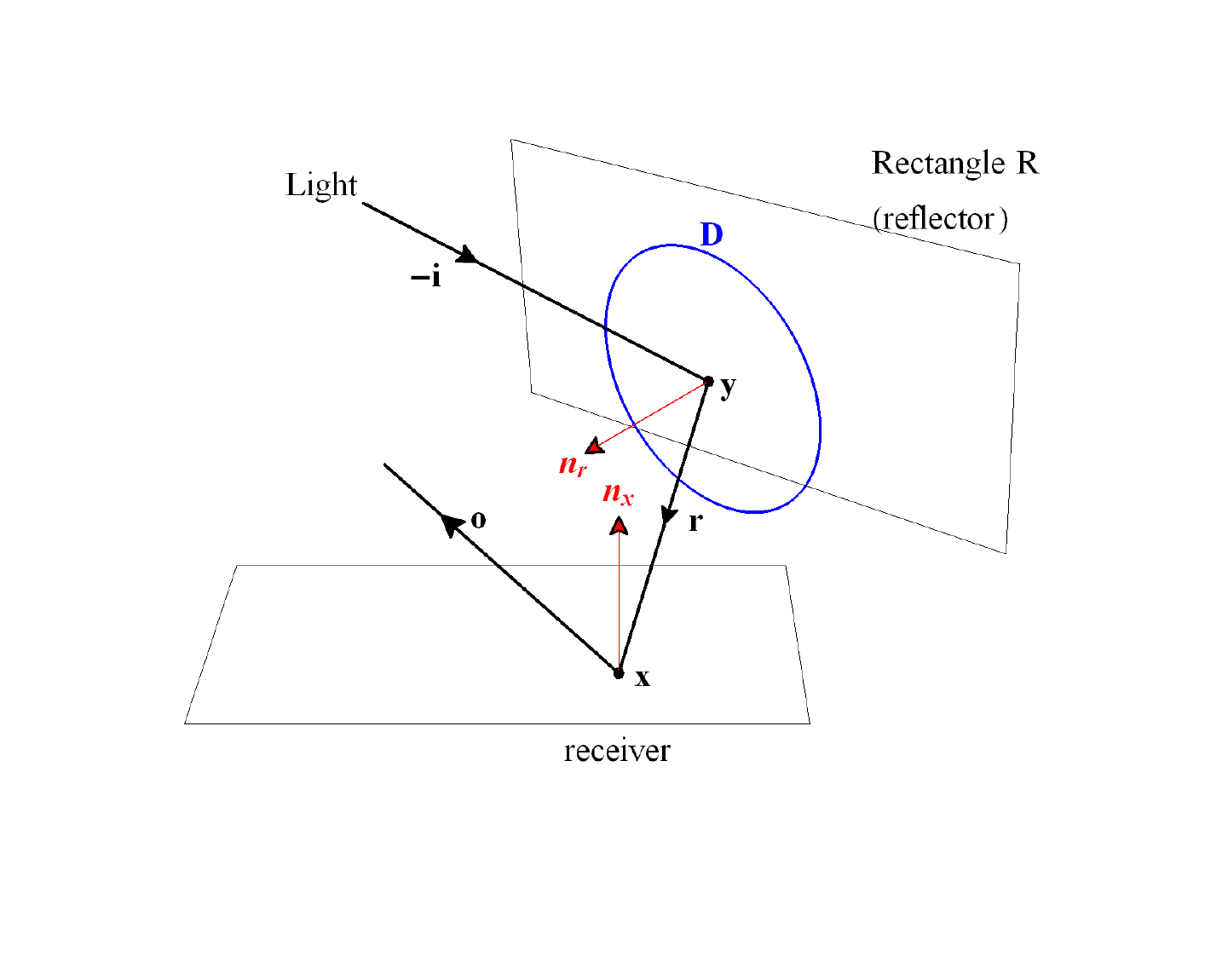}
	\vspace{-6mm}
	\caption{Light path for one-bounce interreflection.}
	\label{fig:findSpecPeak}
\end{figure}

\subsection{Algorithm Overview} 

Our method aims to approximate the glossy interreflections with all-frequency BRDFs and describe the efficient approximation of one-bounce interreflections from a single rectangle (reflector) to a single shading point (receiver). We assume that the lighting is distant and that a single rectangle has a uniform BRDF. We observe that the reflected radiance in a plane (to a shading point) is attenuated around a point (Figure~\ref{fig:overview_a}). We thereafter select a representative point (referred to as the \textit{specular peak}), and a sampling disk (centered at the specular peak) is subsequently created to integrate the reflected radiance. We integrated the reflected radiance as the disk was projected onto a sphere (centered on a shading point)(Figure~\ref{fig:overview_b}). Further, the reflected radiance is approximated with the ASG by preserving the area and function energy (Figure~\ref{fig:overview_c}). Finally, the normal distribution function (NDF) at the receiver is approximated as SG and subsequently convolved with ASG light to compute the final lighting.

\subsection{Rendering Model} 
We start by deriving the one-bounce interreflection model from a single rectangular reflector under distant lighting, as shown in figure~\ref{fig:findSpecPeak}. Given a rectangle $\rectangle$ with normal $\rectNormal$ and a shading point $\recvPt$ with normal $\recvNormal$, we aim to compute the outgoing radiance from $\recvPt$ to the view direction $\recvPtViewVec$ after the reflection of light from the rectangle $\rectangle$ toward $\recvPt$. To simplify the derivation, we assume that there is no occlusion between the light, reflected rectangle, and shading point. The one-bounce radiance from $\recvPt$ toward $\recvPtViewVec$ can be expressed as a standard rendering equation \cite{Kajiya1986}:
\begin{equation}\label{eq:recvRectangleEq}
	L_x(\recvPtViewVec)=\int_{\Omega_R}  L(\peakPtViewVec)\, f_r(-\peakPtViewVec, \recvPtViewVec), (-\peakPtViewVec \cdot \recvNormal)  \D \peakPtViewVec. 
\end{equation}
where $\peakPtViewVec$ denotes the direction from a point on the reflected rectangle $\rectangle$ to $\recvPt$, and the integral is over the rectangle $\Omega_R$ subtended by $\rectangle$; $f_r$ and $\recvNormal$ are the BRDF and normal directions at the receiver point $\recvPt$, respectively, and $L(\peakPtViewVec)$ is the reflected radiance from any point (on rectangle $\rectangle$) to $\recvPt$, defined as follows:
\begin{equation}\label{eq:reflRenderEq}
	L(\peakPtViewVec)=\int_{\Omega}  L(\lightVec)\, f_r(\lightVec, \peakPtViewVec)\, (\lightVec \cdot \rectNormal)  \D \lightVec. 
\end{equation}
where $\rectNormal$ denotes the normal of the rectangle $\rectangle$, and $L(\lightVec)$ is the incident light.

Here, we use a simplified Disney BRDF model \cite{McAuley2012, Karis2013} for $f_r$:
\begin{equation}
	\begin{aligned}
		f_r(\lightdir, \viewdir) &=\mathcal{M}(\lightdir, \viewdir) \, \mathcal{D}(\mathbf{h}),
	\end{aligned}
\end{equation}
where $\lightdir$ denotes the incident light direction, $\viewdir$ is the viewing direction, $\mathbf{h} = (\viewdir + \lightdir) / \|\viewdir + \lightdir\|$ is the half vector, $\mathcal{M}$ is the Fresnel and shadowing effects, and $\mathcal{D}$ is the NDF. 

We represent the NDF at shading point $\recvPt$ as a uniform SG \cite{Wang2009}, which is also a von Mises--Fisher distribution: 
\begin{align}
	\begin{aligned}
		\mathcal{D}(\mathbf{h})& = G_{i}(\mathbf{h}; \recvNormal, \dfrac{2}{\roughness^2}, \dfrac{1}{\pi \roughness^2})
	\end{aligned}
\end{align}
where $\recvNormal$ is the surface normal, and $\roughness$ is the material roughness at the shading point. 

We represent the $L(\peakPtViewVec)$ as an ASG \cite{Xu2013}:
\begin{align}\label{eq:asgLight}
	L(\peakPtViewVec)=G_{a}(-\peakPtViewVec; [\asgLobe,\asgTangent,\asgBitangent], [\asgLambda,\asgMu],\asgAmp)
\end{align}
We denote as $G_{a}(\cdot)$ for notation simplicity. The computation of the ASG parameters is described in Sec.~\ref{secASGLight}.

To convolve with light ASG, we need to express $\mathcal{D}$ in terms of lighting vector rather than half vector. The Jacobian of this change is $4\|\mathbf{h} \cdot \recvPtViewVec\|$, and $\mathcal{D}$ is,
\begin{align}
	\mathcal{D}(\mathbf{h}) = G_{i}(-\peakPtViewVec; \mathbf{r_x}, \dfrac{2}{4\|\mathbf{h} \cdot \recvPtViewVec\|\, \roughness^2}, \dfrac{1}{\pi \roughness^2})
\end{align}
where $\mathbf{r_x}$ denotes the new SG lobe axis, which can be calculated by $\mathbf{r_x}=2\|\recvPtViewVec \cdot \mathbf{h}\|\, \mathbf{h}-\recvPtViewVec$. We denote as $G_{i}(\cdot)$ for notation simplicity.

Now, we can rewrite Eq.~\ref{eq:recvRectangleEq} as follows:
\begin{equation}
	\begin{aligned}
		L_x(\recvPtViewVec) \approx \int_{S^2}  &G_{a}(\cdot) G_{i}(\cdot) \mathcal{M}(-\peakPtViewVec, \recvPtViewVec), (-\peakPtViewVec \cdot \recvNormal) \, \D \peakPtViewVec.
	\end{aligned}
\end{equation}
Since the last two terms are low-frequency terms, they can be approximated by a constant, and we can find a specific location $\peakPt$(explained in Sec.~\ref{secFindPeak}) at rectangle $\rectangle$ and thereafter pull them out of the integral:
\begin{equation}\label{eq:finalRender}
	\begin{aligned}
		L_x(\recvPtViewVec) \approx \mathcal{M}(-\peakPtViewVec_y, \recvPtViewVec) \, (-\peakPtViewVec_y \cdot \recvNormal) \cdot \int_{S^2}  &G_{a}(\cdot) G_{i}(\cdot) \D \peakPtViewVec.
	\end{aligned}
\end{equation}
where $\peakPtViewVec_y$ denotes a direction point from the representative point $\peakPt$ (on the reflected rectangle $\rectangle$) to the shading point $\recvPt$.

The product integrals of $G_{a}(\cdot)$ and $G_{i}(\cdot)$ were derived by Xu \cite{Xu2007}:
\begin{equation}\label{eq:sgConvAsg}
	\begin{aligned}
		\mathbf{H(p)}&= \int_{S^2} G_a(v;[z,x,y],[\lambda,\mu],1) \cdot G_i(v;p,\nu,1) \D \mathbf{v} \\ 
		& \approx G_a(p,[z,x,y],[\dfrac{\nu\lambda}{\nu+\lambda},\dfrac{\nu\mu}{\nu+\mu}],\dfrac{\pi}{\sqrt{(\lambda+\nu)(\mu+\nu)}})
	\end{aligned}
\end{equation}

Now, the problem of calculating $L_x(\recvPtViewVec)$ is reduced to two questions: finding the representative point $\peakPt$ on rectangle $\rectangle$ to approximate the low-frequency terms and determining the parameters of $G_{a}(\cdot)$ to closely approximate $L(\peakPtViewVec)$.

\subsection{Finding Specular Peak} \label{secFindPeak}
As shown in figure.~\ref{fig:findSpecPeak}), for a given rectangle $\rectangle$ centered at $\rectCenter$, with a normal $\rectNormal$, 
the shading point $\recvPt$ is at position $\recvPtPos$, and the distance in the light direction is $\lightVec$. Since we find a point on $\rectangle$ where the normal vector is equal to the half vector, we first compute the reflected direction $\peakPtViewVec$ at point $\peakPt$ by $\peakPtViewVec=2(\lightVec \cdot \rectNormal)\rectNormal-\lightVec$. Thereafter, the position of the peak point $\peakPt$, $\peakPtPos$, can be calculated as follows:
\begin{equation}\label{eq:findSpecPeak}
	\peakPtPos=\recvPtPos -\peakPtViewVec \cdot \dfrac{(\recvPtPos - \rectCenter) \cdot \rectNormal}{\lightVec \cdot \rectNormal}, \quad \lightVec \cdot \rectNormal > 0
\end{equation}
We thereafter used this point to approximate the low-frequency terms in Eq.~\ref{eq:finalRender}

\subsection{Integrating Reflected Radiance} \label{secIntsReflect}
We approximate the ASG light $G_a(\cdot)$ by preserving the area and function energy of the reflected radiance from the rectangle $\rectangle$. In this section, we introduce the approximation of the total reflected radiance from $\rectangle$, and the result is used to create $G_a(\cdot)$ in Sec.\ref{secASGLight}.

The total reflected radiance from rectangle $\rectangle$ to shading point $\recvPt$ is obtained as follows:
\begin{equation}
	\mathbf{C}=\int_{\Omega_R}  L(\peakPtViewVec) \D \peakPtViewVec. 
\end{equation}

To approximate the reflected radiance from rectangle $\rectangle$ to shading point $\recvPt$, we first select a sampling disk $\disk(\peakPt, \diskRadius, \diskNormal)$, with $\peakPt$, $\diskRadius$, and $\diskNormal$ as the center point, radius, and normal of the disk, respectively, where $\peakPt$ is the position of the specular peak point, $\diskNormal$ is equal to the normal of rectangle $\rectNormal$, and $\diskRadius$ is the radius of the disk; $\diskRadius$ is a user-controlled variable. We approximate $\mathbf{C}$ as follows:
\begin{equation}\label{eq:recvRenderEq}
	\mathbf{C} \approx \areaPercent \cdot \int_{\Omega_D}  L(\peakPtViewVec) \D \peakPtViewVec. 
\end{equation}
The integral changes to $\Omega_D$ subtended by $\disk$; $\areaPercent$ is the intersection area divided by the disk area, and it is explained in Sec.\ref{secRectProxy}. 

\medskip
We first evaluate the reflected radiance from specular peak $\peakPt$ to shading point $\recvPt$ because $\peakPt$ is the sampled point on rectangle $\rectangle$, and its value can be estimated using the Monte Carlo method as follows:
\begin{equation}\label{eq:reflMonteEqIter}
	L_y(\peakPtViewVec) \approx \dfrac{1}{N} \sum_{k=1}^{N} \dfrac{{f_r(\mathbf{i_k}, \mathbf{r_k})\, (\mathbf{i_k} \cdot \mathbf{n_y,_k})} }{p(\mathbf{i_k},\mathbf{r_k})}.
\end{equation}
where $N$ denotes the light number in the scene, $\mathbf{i_k}$, $\mathbf{r_k}$, and $\mathbf{n_y,_k}$ correspond to $\lightVec, \peakPtViewVec, and \peakPtNormal$ Eq~{\ref{eq:reflRenderEq}}, related to the $k$-th light. The term $p(\mathbf{i_k},\mathbf{r_k})$ is the probability density function(PDF) with respect to the solid angle for the $k$-th light. For notation simplicity, we subsequently omit the summation $\sum_{k=1}^{N}(\cdot)$ over index $k$. 

\medskip
The term $p(\mathbf{i},\mathbf{r})$ can be calculated using the NDF. In this study, we used the GGX microfacet BRDF \cite{Walter2007} as NDF, which is currently considered the most realistic parametric BRDF \cite{Hill2015}. The definition of GGX is obtained as follows:
\begin{equation}\label{eq:ggx}
	\mathcal{D}(\boldsymbol{\theta}, \roughness) = \dfrac{\roughness^2}{\pi (1-(1-\roughness^2) \cdot \cos(\boldsymbol{\theta})^2)^2}
\end{equation}
where $\roughness$ denotes the material roughness, and $\boldsymbol{\theta}$ is the angle between the normal vector and the half vector. The parameter $\boldsymbol{\theta}$ is 0 at the specular peak $\peakPt$, that is, the normal vector is equal to the half vector.

Next, we substitute $p(\mathbf{i},\mathbf{r})=\mathcal{D}(\theta, \rectRoughness)\cos{\theta}$ with $\theta=0$ in Eq~\ref{eq:reflMonteEqIter}, which yields,
\begin{equation}\label{eq:reflMonteEq}
	L_y(\peakPtViewVec) \approx \pi \rectRoughness^2 f_r(\mathbf{i}, \mathbf{r})\, (\mathbf{i} \cdot \mathbf{n_y}).
\end{equation}
where $\rectRoughness$ denotes the material roughness of rectangle $\rectangle$.

\medskip
After estimating the reflected radiance from $\peakPt$, we integrate the total reflected radiance from the sampling disk $\disk$. 
First, we create a sphere centered at $\recvPt$ with radius $\|\peakPtPos-\recvPtPos\|$ then integrate over a spherical cap to compute the reflected radiance from disk $\disk$ because the disk is projected onto this sphere, as shown in Figure~\ref{fig:overview_b}. For notation simplicity, but without loss of generality, we assume that the distance between $\recvPt$ and $\peakPt$ is 1 in the remainder of this study, that is, we are working on a unit sphere. We observe that the reflected radiance is proportional to the angle between the reflected rectangle normal $\rectNormal$ and shading normal $\recvNormal$. Subsequently, we select a \textit{representative attenuation axis} $\uAxis$ and assume that the radiance attenuation from specular peak $\peakPt$ toward all directions is the same as that in direction $\uAxis$, and it can be calculated as follows (Figure~\ref{fig:overview_a}):
\begin{equation}\label{eq:uvaxis}
	\begin{cases}		
		\vAxis &= \vec{\rm yx} \times \vec{\rectNormal} \\
		\uAxis &= \vec{\rectNormal} \times \vAxis
	\end{cases}
\end{equation}

\begin{figure}[!b]
\includegraphics[trim=0 0 0 100, clip, width=\linewidth]{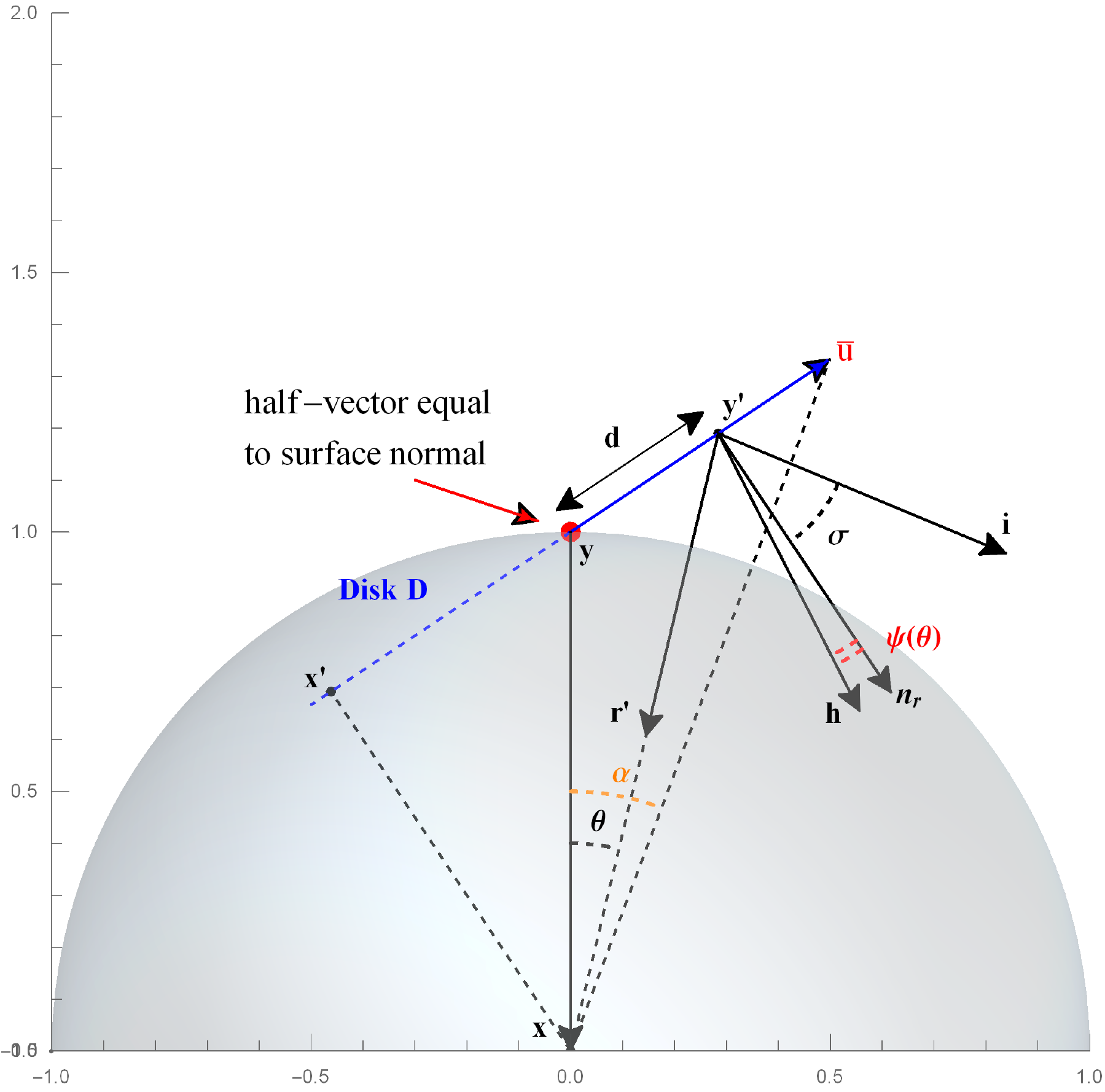}
\vspace{-6mm}
\caption{Solving $\reflNohAngle$.}
\label{fig:diffNoH}
\end{figure}

To demonstrate the derivation of the integral but without loss of generality, we use $\vec{\rm xy}$ as the zenith axis, where the axis $\vAxis$ points toward the screen, as shown in Figure~\ref{fig:diffNoH}. 

Now, we approximate the integral term in Eq.~\ref{eq:recvRenderEq} as follows:
\begin{equation}\label{eq:intsConeA}
	\begin{aligned}			
		\mathbf{E} &= \int_{\Omega_D}  L(\peakPtViewVec) \D \peakPtViewVec \\ 
		&\approx \int_0^{2\pi} \int_0^{\capAperture}  L_c(\theta,\phi)\, \sin{\theta}\, \D \theta\, \D \phi.
	\end{aligned}
\end{equation}
where $L_c(\theta,\phi)$ denotes the reflected radiance in spherical coordinates $(\theta, \phi)$, and $\capAperture$ is the aperture angle of integrating the spherical cap (as shown in Figure~.\ref{fig:diffNoH}); the calculation of $\capAperture$ will be explained later. Since we already assume that the attenuation of the reflected radiance is represented by axis $\uAxis$, we can directly integrate the $\phi$ term as follows:
\begin{equation}\label{eq:intsConeA2}
	\begin{aligned}			
		\mathbf{E} \approx 2\pi \int_0^{\capAperture} L_c(\theta)\, \sin{\theta}\, \D \theta
	\end{aligned}
\end{equation}

$L_c(\theta)$ can be estimated in the same manner as Eq. ~\ref{eq:reflMonteEq}:
\begin{equation}\label{eq:intsConeA3}
	\begin{aligned}	
		L_c(\theta) &\approx \pi \rectRoughness^2 f_r(\lightVec, \peakPtViewVec)\, (\lightVec \cdot \mathbf{n_y}) \\
		&= \pi \rectRoughness^2 \mathcal{M}(\mathbf{i}, \peakPtViewVec) \cdot \mathcal{D}(\reflNohAngle,  \rectRoughness)\, (\lightVec \cdot \mathbf{n_y}).
	\end{aligned}
\end{equation}
where $\reflNohAngle$ denotes an angle function that represents the angle between the normal and half vectors at a differential point (e.g.,  $\mathbf{y'}$ on the reflector disk in Figure~\ref{fig:diffNoH}), and its value depends on the $\theta$ in the integrator. Notably, $L_c(0)=L_y(\peakPtViewVec)$. 

Since $\mathcal{M}$ is a low-frequency term, we use its value at point $\peakPt$ and pull the low-frequency terms out of the integral as follows:
\begin{equation}\label{eq:intsDisk}
	\begin{aligned}		
		\mathbf{E} &= 2\pi \int_0^{\capAperture} \pi \rectRoughness^2\, \mathcal{M}(\mathbf{i}, \peakPtViewVec) \, \mathcal{D}(\reflNohAngle, \rectRoughness)\, (\mathbf{i} \cdot \mathbf{n_y})\, \sin{\theta}\, \D \theta \\
		&\approx 2\pi\, \cdot \pi\, \rectRoughness^2\, \mathcal{M}(\mathbf{i}, \peakPtViewVec)\,  (\mathbf{i} \cdot \mathbf{n_y})\,  \int_0^{\capAperture} \mathcal{D}(\reflNohAngle, \rectRoughness)\, \sin{\theta}\, \D \theta \\
		&= 2\pi\, L_y(\peakPtViewVec)\, \int_0^{\capAperture} \dfrac{\mathcal{D}(\reflNohAngle, \rectRoughness)}{\mathcal{D}(0, \rectRoughness)}\, \sin{\theta}\, \D \theta, \\
		&=  2\pi\, L_y(\peakPtViewVec)\, \int_0^{\capAperture}\, f(\theta)\, \sin{\theta}\, \D \theta
	\end{aligned}
\end{equation}
We denote $f(\theta)=\dfrac{\mathcal{D}(\reflNohAngle, \rectRoughness)}{\mathcal{D}(0, \rectRoughness)}$ and substitute Eq. ~.\ref{eq:ggx} into $f(\theta)$:
\begin{equation}\label{eq:integrandA}
	f(\theta) = \dfrac{\rectRoughness^4}{(1-(1-\rectRoughness^2) \cdot \cos^2(\reflNohAngle)\,)^2}
\end{equation}

\medskip
The solution of $\reflNohAngle$ is shown in Figure~\ref{fig:diffNoH}. For a differential angle $\theta$, we can find a point $\mathbf{y'}$ on the reflector disk $\disk$ along the representative axis $\uAxis$. We denote $\mathbf{\sigma}$ as the angle between the light vector $\lightVec$ and the reflector normal vector $\rectNormal$, $\mathbf{h}$ is the half vector at point $\mathbf{y'}$, and $\mathbf{d}$ is the distance between $\mathbf{y'}$ and $\peakPt$. Thereafter, $\mathbf{d}$ can be approximated as $\mathbf{d} \approx \dfrac{\tan{\theta}}{\cos{\sigma}}$. Hence,
\begin{equation}\label{eq:diffNohAngle}
	\reflNohAngle \approx -\dfrac{\sigma}{2}\, \, + \dfrac{1}{2}\, \arctan(\sec^2{\sigma}\, \tan{\theta} + \tan{\sigma}\,)
\end{equation}
Unfortunately, the integral $\mathbf{E}$ does not have an analytic solution. To solve Eq.~\ref{eq:intsDisk}, we first fit $f(\theta)$ to the following equation:
\begin{equation}\label{eq:fitIntegrand}
	\begin{split}
		f(\theta) &\approx e^{-k \cdot \sin^2{\theta}}\, \cos{\theta} ,\quad k=\dfrac{0.288\, \cos{\sigma}}{\rectRoughness^2}-0.673
	\end{split}	
\end{equation}

\begin{figure}
	\centering
	\begin{subfigure}{0.45\columnwidth}
		\centering
		\includegraphics[width=\textwidth]{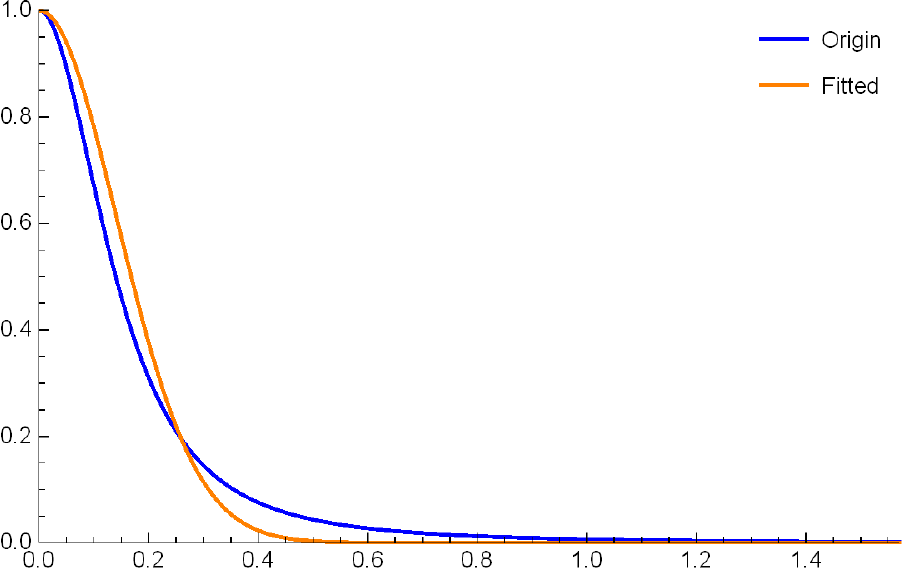}
		\caption{$\sigma=\dfrac{\pi}{6}, \rectRoughness=0.1$}
		\label{fig:compareFit_a}
	\end{subfigure}
	\hfill
	\begin{subfigure}{0.45\columnwidth}
		\centering
		\includegraphics[width=\textwidth]{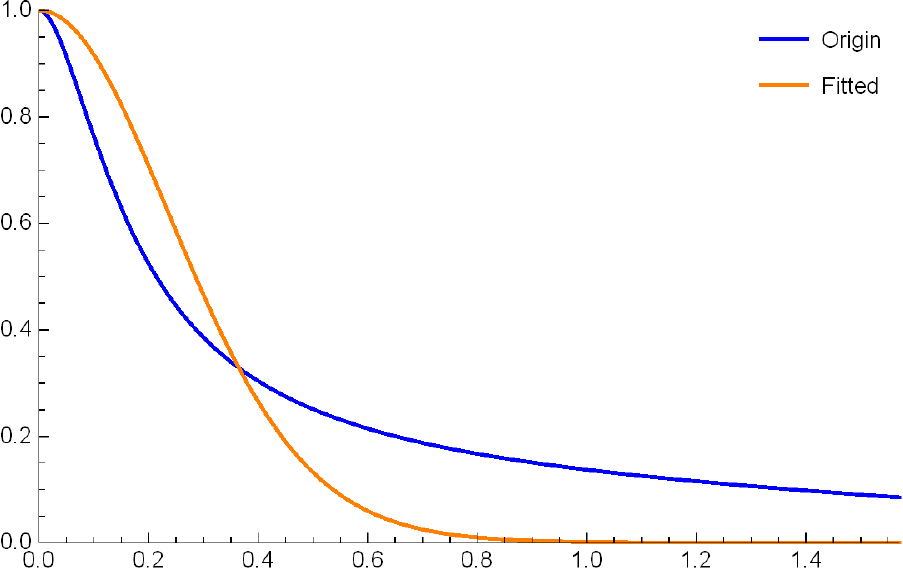}
		\caption{$\sigma=\dfrac{2\pi}{5}, \rectRoughness=0.1$}
		\label{fig:compareFit_b}
	\end{subfigure}
	\vskip\baselineskip
	\begin{subfigure}{0.45\columnwidth}
		\centering
		\includegraphics[width=\textwidth]{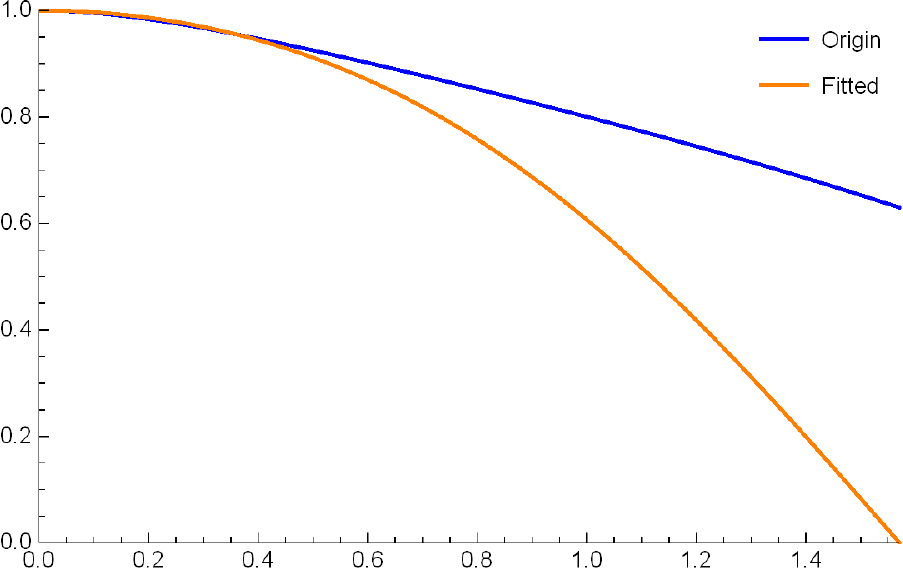}
		\caption{$\sigma=\dfrac{\pi}{6}, \rectRoughness=0.7$}
		\label{fig:compareFit_c}
	\end{subfigure}
	\hfill
	\begin{subfigure}{0.45\columnwidth}
		\centering
		\includegraphics[width=\textwidth]{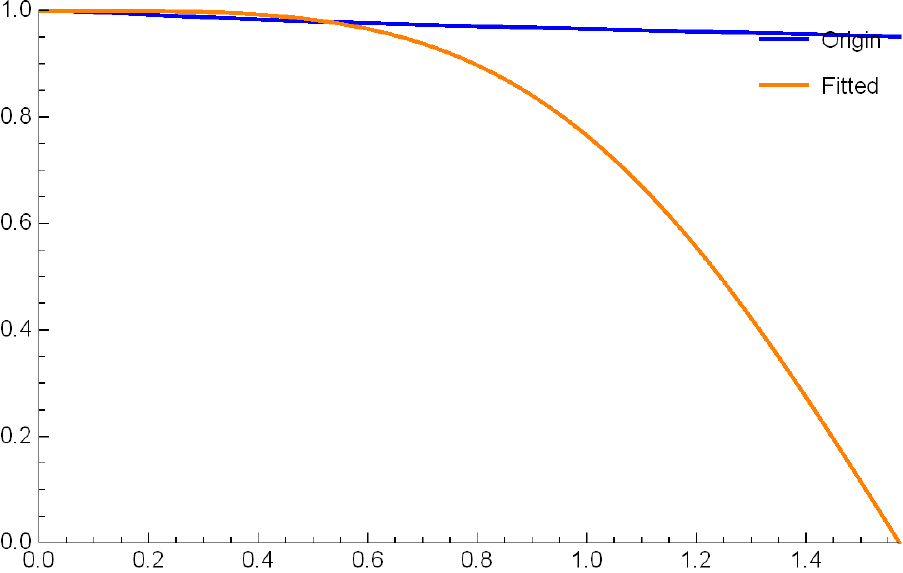}
		\caption{$\sigma=\dfrac{2\pi}{5}, \rectRoughness=0.7$}
		\label{fig:compareFit_d}
	\end{subfigure}
	\caption{Fit error of Eq.~\ref{eq:diffNohAngle}}:
	\label{fig:compareFit}
\end{figure}

The fit error is shown in Figure~\ref{fig:compareFit}, and the fitted shape(Eq.~\ref{eq:fitIntegrand}) is close to the original curve shape (Eq.~\ref{eq:integrandA}) when $\sigma$ and $\rectRoughness$ are both small(Figure~\ref{fig:compareFit_a}), but will produce large errors when one of them is large(Figure~\ref{fig:compareFit_b}, Figure~\ref{fig:compareFit_c}, Figure~\ref{fig:compareFit_d}). However, in practice, when $\sigma$ or $\rectRoughness$ is large, the value of $f(\theta)$ is small and contributes less to the entire reflected radiance. Hence, the approximation error in this case does not significantly affect the result. 

Substituting Eq.~\ref{eq:fitIntegrand} into Eq.~\ref{eq:intsDisk}, we obtain the following:
\begin{equation}
	\begin{aligned}	
		\mathbf{E} &\approx 2\pi\, L_y(\peakPtViewVec)\, \int_0^{\capAperture}\, e^{-k \cdot \sin^2{\theta}}\, \cos{\theta}\, \sin{\theta}\, \D \theta \\ 
		&= 2\pi\, L_y(\peakPtViewVec) \cdot (-\dfrac{1}{2k}\, e^{-k \cdot \sin^2{\theta}} \Biggr|_{0}^{\capAperture}). \\
		&= \dfrac{\pi}{k}\, L_y(\peakPtViewVec) \cdot (1-e^{-k \cdot \sin^2{\capAperture}})
	\end{aligned}
\end{equation}	
Thereafter, we approximate $\capAperture$, the upper limit of the integral, as $\capAperture \approx \arctan(\diskRadius \cdot \cos{\sigma})$, where $\diskRadius$ denotes the radius of the sampling disk $\disk$; then, $\sin^2{\capAperture} \approx 1-\dfrac{1}{1+\diskRadius^2\,\cos^2{\sigma}}$. The radius of the sampling disk is a user control variate, and the choice of $\diskRadius$ will be discussed in Sec.~\ref{sec:result}).

\medskip
In conclusion, the total reflected radiance from the sampling disk $\disk$ to the shading point $\recvPt$ is approximated by the following equation:
\begin{equation} \label{eq:intsRadiance}
	\mathbf{C} \approx \areaPercent \cdot \dfrac{\pi}{k}\, (1-e^{-k \cdot t})\, L_y(\peakPtViewVec)
\end{equation}	
with $k=\dfrac{0.288\, \cos{\sigma}}{\rectRoughness^2}-0.673$ and $t=1-\dfrac{1}{1+\diskRadius^2\,\cos^2{\sigma}}$. 

\medskip
Since $\cos{\sigma}=\rectNormal \cdot \lightVec$, this approximation has a performance advantage in avoiding costly inverse trigonometry operations on the GPU.

\subsection{Approximating ASG light}\label{secASGLight}
We compute the parameters of the ASG light introduced in Eq.\ref{eq:asgLight}. The shape of the projected area is similar to that of an ellipse, as shown in Figure~\ref{fig:overview_b}. We subsequently approximate the reflected radiance as an ASG light(Figure~\ref{fig:overview_c}), which is denoted as $G_{a}(\mathbf{v}; [\asgLobe,\asgTangent,\asgBitangent], [\asgLambda,\asgMu],\asgAmp)$. The lobe axis $\asgLobe$ can be directly calculated using $\asgLobe=\dfrac{\peakPtPos-\recvPtPos}{\| \peakPtPos-\recvPtPos \|}$.

\medskip
The tangent axis $\asgTangent$ and bitangent axis $\asgBitangent$ correspond to the $\vAxis$ and $\uAxis$ axes of the sampling disk $\disk$, respectively. As shown in Figure~\ref{fig:overview_b}, $\asgTangent$ and $\asgBitangent$ can be solved using the following equation:
\begin{equation}\label{eq:asgTangents}
	\begin{cases}		
		\asgTangent &= \vAxis \\
		\asgBitangent &= \vec{\rm yx} \times \vAxis
	\end{cases}
\end{equation}

We denote $\asgTaSize$, $\asgBiSize$ as the length of $\asgTangent$, $\asgBitangent$; then,
\begin{equation}\label{eq:asgAxisLength}
	\begin{cases}		
		\asgTaSize &= \diskRadius \\
		\asgBiSize &= \diskRadius \, (\diskNormal \cdot \vec{\rm yx})
	\end{cases}
\end{equation}

Before solving the bandwidth, we first introduce the polar version of $G_a$, which is equal to Eq.~\ref{eq:asg}:
\begin{equation}\label{eq:asgPolar}
	G_p([\theta, \phi, \eta], [\lambda,\mu], c) = c \cdot \cos{\theta} \cdot e^{-\lambda \cos^2{\phi}-\mu \cos^2{\eta}}.
\end{equation}
Compared to Eq.~\ref{eq:asg}, $\theta$ denotes the polar angle between vector $\mathbf{v}$ and $\mathbf{z}$, $\phi$ is the polar angle between vector $\mathbf{v}$ and $\mathbf{x}$, $\eta$ is the polar angle between vectors $\mathbf{v}$ and $\mathbf{y}$, and the definitions of $\lambda$, $\mu$, and $c$ remain the same.

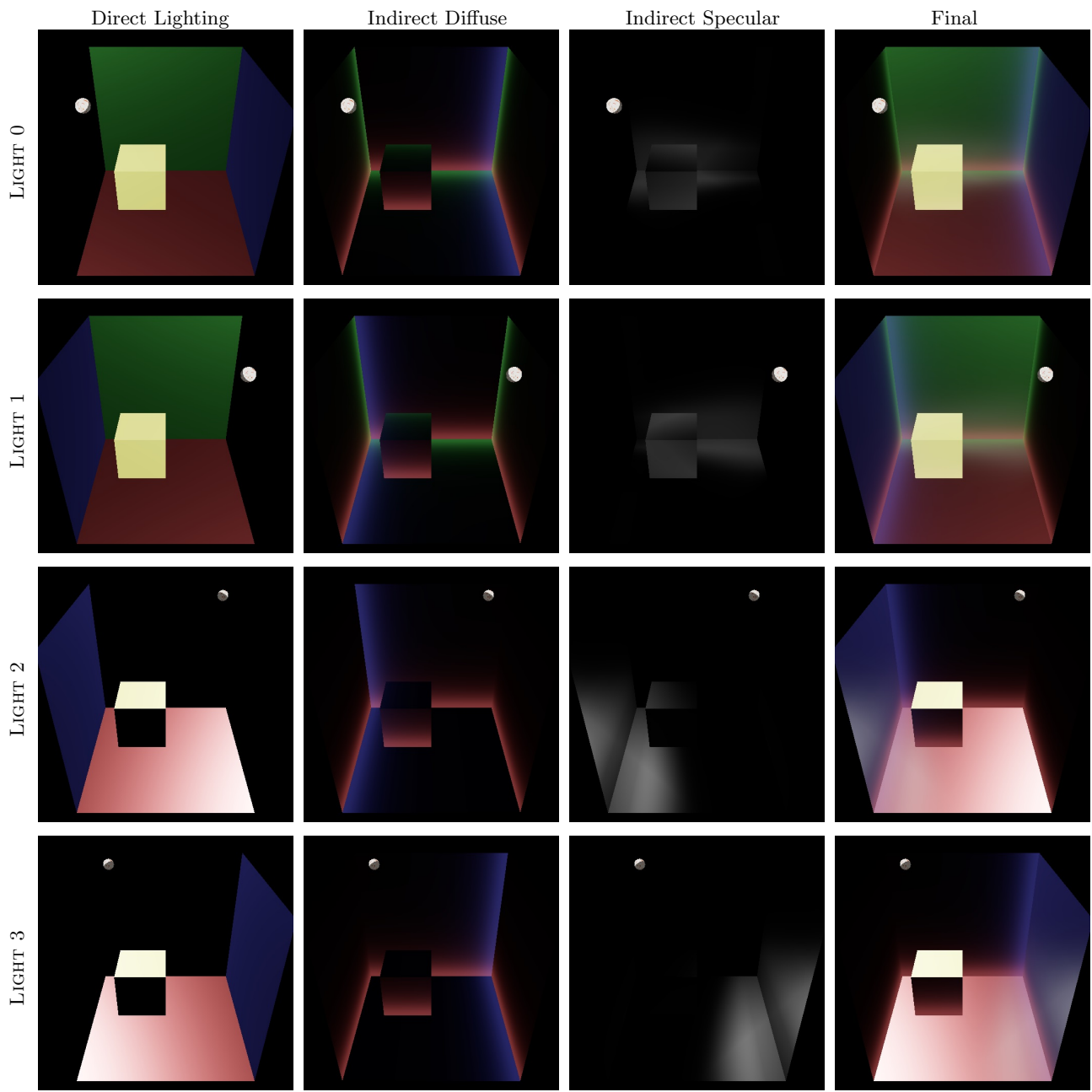
\begin{figure*}[htp]
    \centering
    \setlength{\fboxsep}{0pt}%
    \setlength{\fboxrule}{0.5pt}%
    \contourlength{0.05em}%
    \vspace*{-2.2\baselineskip}%
    \hspace*{-2.0ex}%
    \begin{tikzpicture}[x=0.24\textwidth, y=0.115\textwidth,every text node part/.style={align=center}]
    	\node[anchor=north west] at (0,  0) {\includegraphics[width=0.23\textwidth]{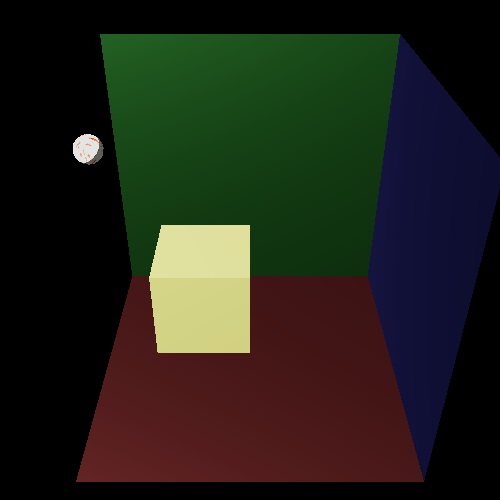}};
    	\node[anchor=north west] at (1,       0) {\includegraphics[width=0.23\textwidth]{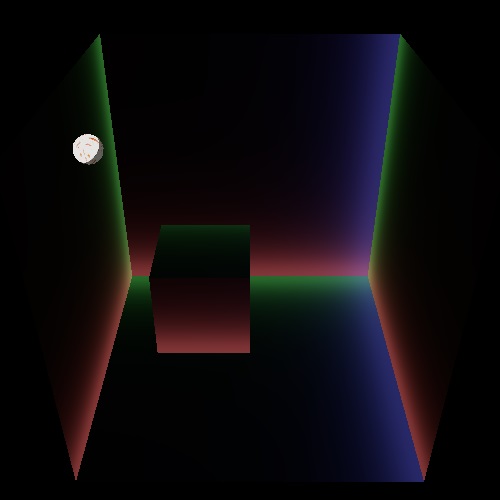}};
    	\node[anchor=north west] at (2,       0) {\includegraphics[width=0.23\textwidth]{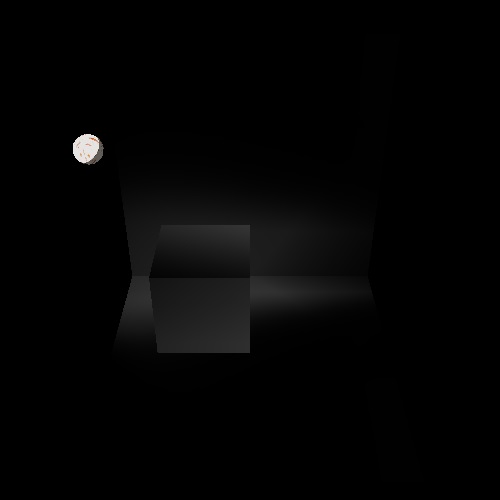}};
    	\node[anchor=north west] at (3,       0) {\includegraphics[width=0.23\textwidth]{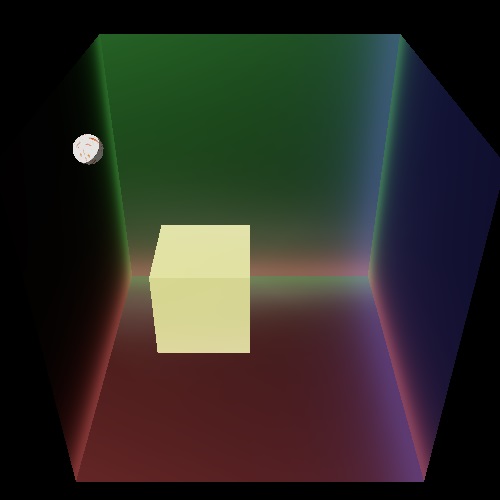}};
    	\node[anchor=north west] at (0,  0- 2.11) {\includegraphics[width=0.23\textwidth]{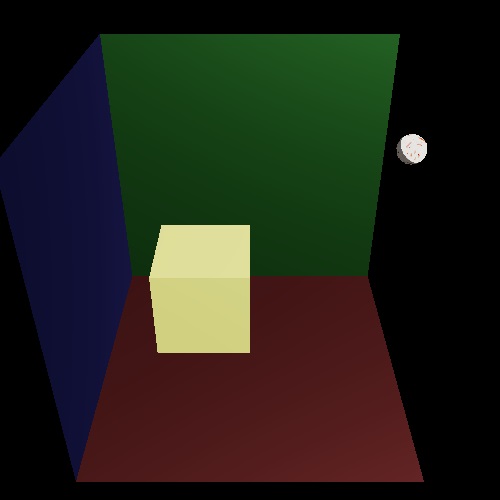}};
    	\node[anchor=north west] at (1,  0- 2.11) {\includegraphics[width=0.23\textwidth]{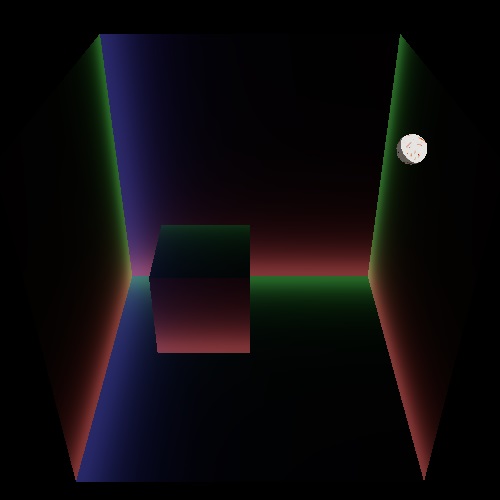}};
    	\node[anchor=north west] at (2,  0- 2.11) {\includegraphics[width=0.23\textwidth]{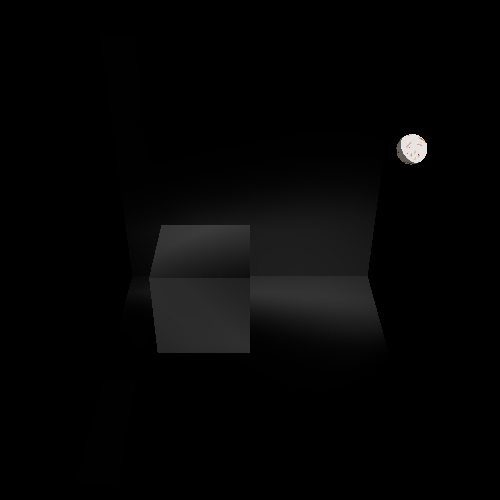}};
    	\node[anchor=north west] at (3,  0- 2.11) {\includegraphics[width=0.23\textwidth]{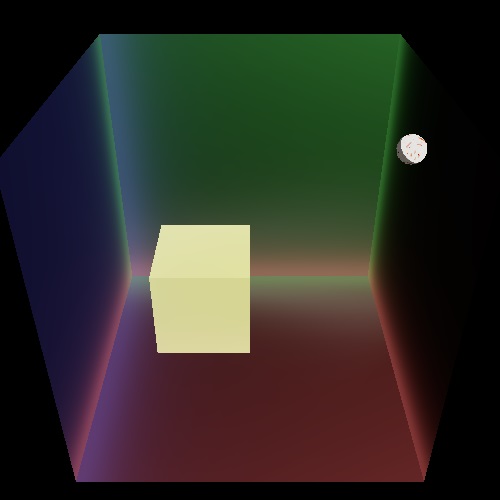}};
    	\node[anchor=north west] at (0,  0-2.11*2) {\includegraphics[width=0.23\textwidth]{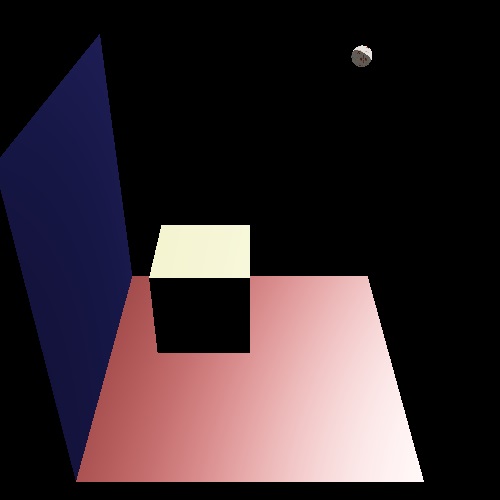}};
    	\node[anchor=north west] at (1,  0-2.11*2) {\includegraphics[width=0.23\textwidth]{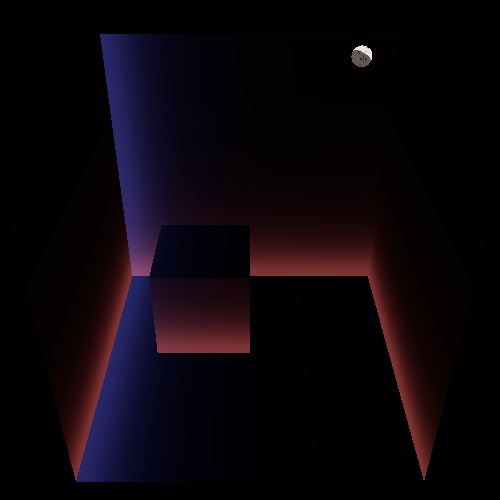}};
    	\node[anchor=north west] at (2,  0-2.11*2) {\includegraphics[width=0.23\textwidth]{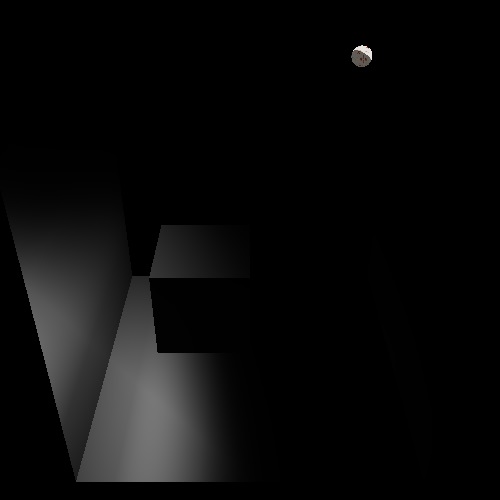}};
    	\node[anchor=north west] at (3,  0-2.11*2) {\includegraphics[width=0.23\textwidth]{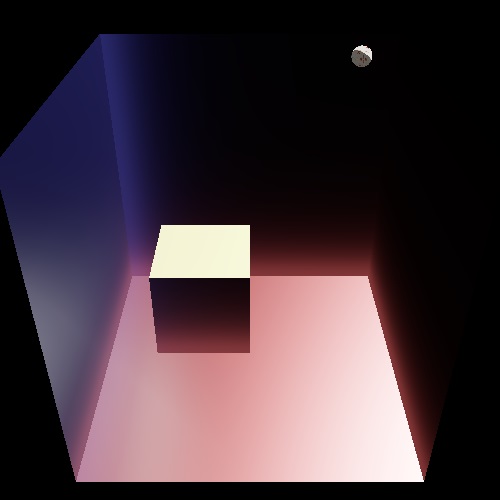}};
    	\node[anchor=north west] at (0,  0-2.11*3) {\includegraphics[width=0.23\textwidth]{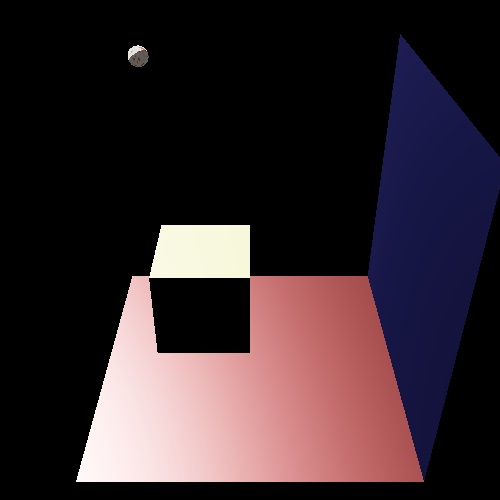}};
    	\node[anchor=north west] at (1,  0-2.11*3) {\includegraphics[width=0.23\textwidth]{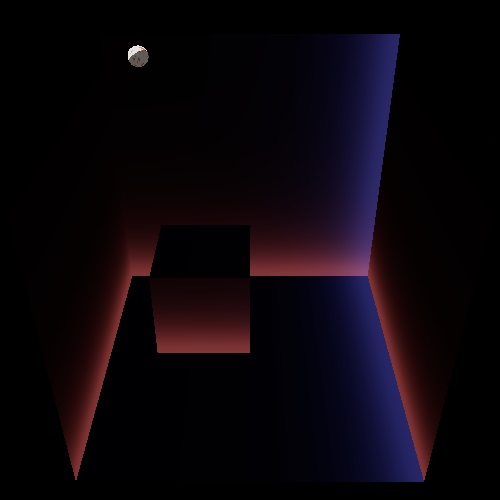}};
    	\node[anchor=north west] at (2,  0-2.11*3) {\includegraphics[width=0.23\textwidth]{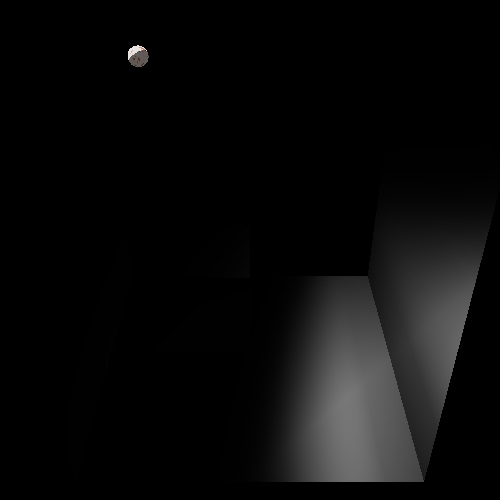}};
    	\node[anchor=north west] at (3,  0-2.11*3) {\includegraphics[width=0.23\textwidth]{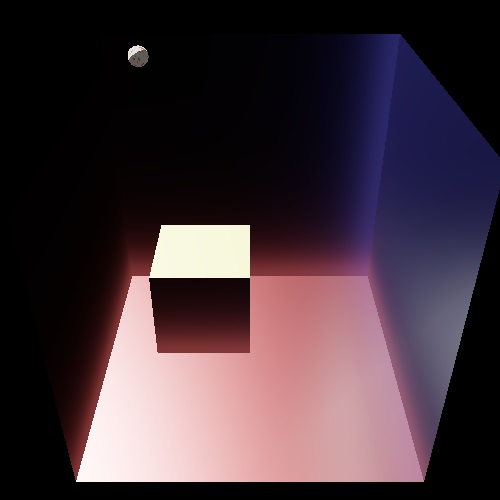}};
    	
    	\node [rotate=90] at (-0.05, -1.1)   {\textsc{\small Light 0}};
    	\node [rotate=90] at (-0.05, -1.1-2.11)   {\textsc{\small Light 1}};
    	\node [rotate=90] at (-0.05, -1.1-2.11*2)   {\textsc{\small Light 2}};
    	\node [rotate=90] at (-0.05, -1.1-2.11*3)   {\textsc{\small Light 3}};
    	
    	\node[anchor=north west] at (0.2,   0.15) {\small Direct Lighting};
    	\node[anchor=north west] at (1.24,  0.15) {\small Indirect Diffuse};
    	\node[anchor=north west] at (2.21,  0.15) {\small Indirect Specular};
    	\node[anchor=north west] at (3.36,  0.15) {\small Final};
    \end{tikzpicture}%
    \caption{Rendering result with different lighting directions.}
    \label{fig:results}
    \vspace{-1.0\baselineskip}
\end{figure*}

\begin{figure}[!b]
	\includegraphics[trim=0 0 0 10, clip, width=\linewidth]{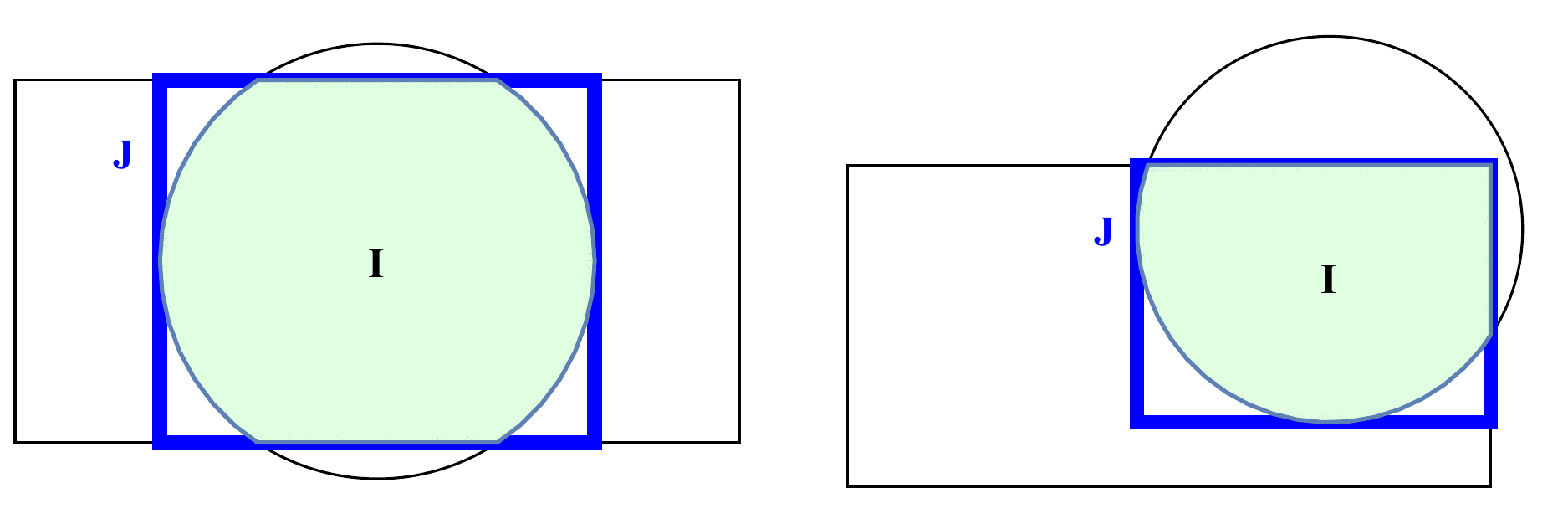}
	\vspace{-6mm}
	\caption{Approximating intersection area.}
	\medskip 
	\small
	The intersection area (green) of disk and rectangle is approximated as a rectangle (blue).
	\label{fig:approxInts}
\end{figure}

\begin{figure}[!b]
	\includegraphics[width=\linewidth]{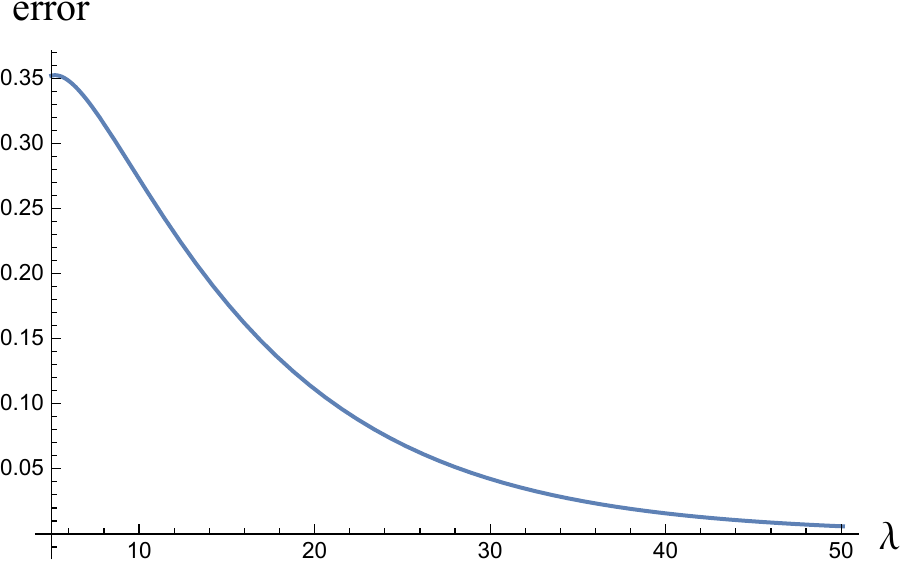}
	\vspace{-6mm}
	\caption{Convolution error.}
	\label{fig:convError}
	\medskip 
	\small
	The error caused by the product integral of the ASG and SG when the bandwidth is small. We show the error in Eq.~\ref{eq:sgConvAsg} with $G_a(v;[(0,0,1),(1,0,0),(0,1,0)],[\lambda,\lambda],1)$, and $G_i(v;(0.5,0,1),2\lambda,1)$ when $\lambda \leq 50$.
\end{figure}

Since ASGs are compactly $\solerror$-supported, we can ensure that in the closed area, $G_{a}$ attains values larger than the threshold $\solerror$, and subsequently obtains the bandwidth $\asgLambda$, $\asgMu$ from shape size $\asgTaSize$, $\asgBiSize$: 
\begin{equation}
	\begin{cases}		
		G_p(\arctan(\asgTaSize),\dfrac{\pi}{2}-\arctan(\asgTaSize),\dfrac{\pi}{2},\asgLambda,\asgMu,1)=\solerror\\ \\
		G_p(\arctan(\asgBiSize),\dfrac{\pi}{2},\dfrac{\pi}{2}-\arctan(\asgBiSize),\asgLambda,\asgMu,1)=\solerror
	\end{cases}
\end{equation}

\medskip
Particularly, we use $\solerror = 0.05$ to calculate the bandwidth:
\begin{equation}\label{eq:asgBandwidth}
	\begin{cases}		
		\asgLambda=(1+\dfrac{1}{\asgTaSize^2})(2.996-0.5\ln(1+\asgTaSize^2)) \\ \\
		\asgMu=(1+\dfrac{1}{\asgBiSize^2})(2.996-0.5\ln(1+\asgBiSize^2))
	\end{cases}
\end{equation}

We obtained the amplitude by preserving the function energy in Eq.~\ref{eq:intsRadiance}, as follows:
\begin{equation}
	\int_{\Omega} G_{a}(\mathbf{v}; [\asgLobe,\asgTangent,\asgBitangent], [\asgLambda,\asgMu],\asgAmp) \D \mathbf{v} = \mathbf{C} 
\end{equation}
The approximation of the ASG integral is as follows:
\begin{equation}
	\int_{\Omega} G_{a}(\mathbf{v}; [\asgLobe,\asgTangent,\asgBitangent], [\asgLambda,\asgMu],\asgAmp) \D \mathbf{v} \approx \dfrac{\pi}{\sqrt{\asgLambda \asgMu}} \, \asgAmp
\end{equation}
then:
\begin{equation}
	\asgAmp \approx \dfrac{\sqrt{\asgLambda \asgMu}}{\pi} \, \mathbf{C}
\end{equation}

\subsection{Rectangle Proxy} \label{secRectProxy}
Our algorithm relies on a \textit{rectangle proxy (RP)} constructed upon each triangle geometry, which closely approximates the geometry surface. For a given scene, the RPs are generated by offline geometry tools for each geometry object. Thereafter, all the proxies are organized into an acceleration structure such as kdtree. We subsequently create one or more light-field textures for each geometry; each texel of light-field texture stores the indices of possible reflected RPs. The indices of static RPs are one-off cache in the related light-field texture, whereas the indices of dynamic RPs are updated to the related light-field texture in each frame. To handle scenes with greater geometric complexity, we can partition the scene, and represent the subspaces with separate and larger RPs that can reduce the RP number and evaluate more rapidly. Because of its compactness and ease of evaluation, the RP model enables real-time rendering with full global illumination effects, particularly for high-frequency glossy interreflections. Since we represent the scene with rectangles, we can calculate the one-bounce radiance from a rectangle to a shading point.

Using a rectangle, we can quickly approximate the intersection area of the rectangle and disk, as shown in figure~\ref{fig:approxInts}. We approximate $\areaPercent$ in Eq.\ref{eq:recvRenderEq} with the following function:
\begin{equation}\label{eq:areaPercentEq}
	\begin{aligned}	
		\areaPercent\, =\, \dfrac{\mathbf{S_I}}{\pi \diskRadius^2}\, \approx\, \dfrac{\mathbf{S_J}}{\pi \diskRadius^2}
	\end{aligned}	
\end{equation}
where $\diskRadius$ denotes the radius of the disk, $\pi \diskRadius^2$ is the area of the disk, $I$ is the real intersection area of the rectangle and disk, $S_I$ is the area of region $I$, $J$ is the axis-aligned bounding rectangle of $I$, and $S_J$ is the area of $J$.

\section{Results and Discussions}

This section presents the results of our algorithm. We present our rendering results on the Cornell Box and the effect when changed to apply distant light to the scene. The error caused by the convolution of the ASG and SG is also discussed.

Figure \ref{fig:results} shows our rendering result with different lighting directions.

In addition, the approximation of Eq.~\ref{eq:sgConvAsg} has a large error when the bandwidth of the ASG is smaller than 50, as shown in Figure \ref{fig:convError}. 

We can approximate the ASG light with an SG to improve performance, which removes the anisotripic property but reduces the calculation cost. The ASG can be approximated as follows:
\begin{equation}
	\begin{aligned}
		G_a(v;[z,x,y],[\lambda,\mu],1) &\approx G_i(v;z,2\lambda,1)
	\end{aligned}
\end{equation}
We assume that $\lambda \geq \mu$, Eq.~\ref{eq:sgConvAsg} can be approximated by the product integral of two SGs, and the calculation of the product integral of two SGs was derived in the work by Iwasaki \cite{Iwasaki2012a}.
 \label{sec:result}

{\small
\bibliographystyle{ieee_fullname}
\bibliography{egbib}
}

\end{document}